\documentclass[aps,pra,amsmath,amssymb,reprint,showpacs]{revtex4-1}
\usepackage{amsmath}
\usepackage{graphicx}
\usepackage{dcolumn}
\usepackage{bm}

\DeclareMathOperator\erfc{erfc}

\begin{document}

\title{Continuous beam of laser-cooled Yb atoms}
\author{K. D. Rathod}
\affiliation{Department of Physics, Indian Institute of
 Science, Bangalore 560\,012, India}
\author{Alok K. Singh}
\affiliation{Department of Physics, Indian Institute of
 Science, Bangalore 560\,012, India}
\author{Vasant Natarajan}
 \affiliation{Department of Physics, Indian Institute of
 Science, Bangalore 560\,012, India}
 \email{vasant@physics.iisc.ernet.in}
 \homepage{www.physics.iisc.ernet.in/~vasant}

\begin{abstract}
We demonstrate launching of laser-cooled Yb atoms in a continuous atomic beam. The continuous cold beam has significant advantages over the more-common pulsed fountain, which was also demonstrated by us recently. The cold beam is formed in the following steps---(i) Atoms from a thermal beam are first Zeeman slowed to a small final velocity, (ii) the slowed atoms are captured in a two-dimensional magneto-optic trap (2D-MOT), and (iii) atoms are launched {\em continuously} in the vertical direction using two sets of moving-molasses beams, inclined at $\pm 15^\circ$ to the vertical. The cooling transition used is the strongly-allowed ${^1S}_0 \rightarrow {^1P}_1$ transition at 399~nm. We capture about $7 \times 10^6$ atoms in the 2D-MOT, and then launch them with a vertical velocity of 13~m/s at a longitudinal temperature of 125(6)~mK.
\end{abstract}



\maketitle

\section{Introduction}
Laser-cooled Yb is an important atom for fundamental studies because the spin-zero ground state obviates the need for a repumping laser, as is required for laser cooling of the more common alkali-metal atoms. Yb has seven stable isotopes, of which five are bosonic and two are fermionic. This allows the
comparative study of Fermi-Bose gas mixtures, particularly
under conditions of quantum degeneracy \cite{TMK03,FTK07}. Degenerate gases of the fermionic isotopes are particularly attractive for the study of synthetic gauge fields \cite{LCJ09,SHV13}. In addition, spin-exchange collisions in the closed-shell
ground state are smaller compared to the alkali-metal
atoms. This makes laser-cooled Yb an attractive candidate
for precision measurements and atomic clocks.
One of us (VN) has previously proposed \cite{NAT05} using
cold Yb atoms launched in an atomic fountain for a  high-precision test of a permanent electric dipole moment (EDM) in an atom. The existence of an atomic EDM would be direct evidence of time-reversal symmetry
violation in the laws of physics. Therefore, EDM searches
are among the most important atomic physics experiments today and can strongly constrain theories that go beyond the
Standard Model. Yb also has two narrow optical transitions that are good candidates for atomic clocks: the $^1{S_0} \rightarrow {^3P}_2$ transition at 507 nm \cite{HZB89}, and the $^1{S_0} \rightarrow {^3P}_0$ transition at 578 nm \cite{HBO05}. Both these transitions can be accessed in atomic fountains.

Here, we demonstrate a continuous (cw) beam of cold Yb atoms, based on a configuration similar to what has been demonstrated earlier in Cs \cite{WAV97}. A cw beam has significant advantages over the more-common pulsed fountain, which was also demonstrated for Yb by us recently \cite{PRS10}. For EDM experiments with a continuous beam, the electric-field plates can be brought very close without worrying about the launching beams, because the launch beams are off-axis from the field plates. For optical clocks, a continuous beam avoids limitations due to intermodulation effects, also known as the Dick effect \cite{GDT07}. For getting the cw cold beam, Yb atoms emanating from an oven are first laser cooled and captured in a two-dimensional magneto-optic trap (2D-MOT), and then launched {\em continuously} using a set of moving-molasses beams inclined at $\pm 15^\circ$ to the vertical. About $7 \times 10^6$ atoms captured in the 2D-MOT are launched with a vertical velocity of 13~m/s at a longitudinal temperature of 125(6)~mK.

\section{Experimental details}

Yb has two cooling transitions---the strongly-allowed $^1{S_0} \rightarrow {^1P}_1$ transition at 399~nm, and the weakly-allowed $^1{S_0} \rightarrow {^3P}_1$  intercombination line at 556~nm. In this study, we have only used the former one. Though its relatively large linewidth of 28~MHz implies a large Doppler-cooling temperature of 690~$\mu$K, it allows for Zeeman slowing over a short distance, and a higher capture velocity in the MOT \cite{PRS10}.

The laser for accessing this transition is generated in a two-step process. We start with a single-frequency Ti:sapphire laser (Coherent 899-21) operating at 798 nm, pumped by 10 W of 532 nm light. The output of the Ti:Sa laser is typically 1.3 W, and it is stabilized on a reference cavity to give an rms linewidth of 1 MHz. This output is then frequency doubled to 399 nm in an external delta-cavity doubler (Laser Analytical Systems), to give an output power of about 160 mW. Of this, 30--40~mW is sent through an acousto-optic modulator (AOM) for the Zeeman-slowing beam.
The remaining power is used to produce the 2D-MOT and launching beams. The required frequency shifts are produced using further AOMs along the beam paths. The launched atoms are probed using a second low-power laser, composed of a grating-stabilized diode laser (Nichia Corporation) with a total output power of 5~mW.

\begin{figure}
\centerline{\resizebox{0.95\columnwidth}{!}{\includegraphics{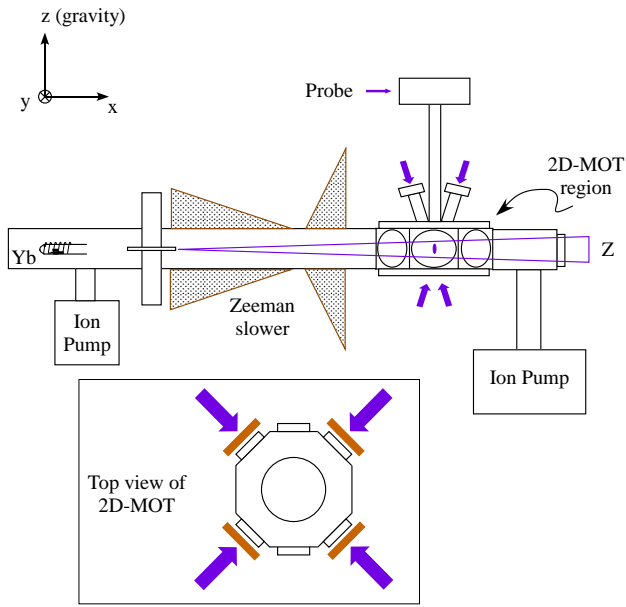}}}
\caption{(Color online) Schematic of the vacuum chamber used for the experiment. Detailed description in the text.}
 \label{schema}
\end{figure}

A schematic of the vacuum system used in the experiment is shown in Fig.\ \ref{schema}. The source of atoms is a resistively-heated quartz ampoule containing all the isotopes of Yb in their natural abundance. The source region is maintained at a pressure below $10^{-7}$ torr using a 20-l/s ion pump. This region is attached to the experimental chamber through a differential-pumping tube, so as to allow for a pressure difference of 2 orders of magnitude. The first part of the chamber is a Zeeman-slowing region, consisting of a tube with 42-mm OD and 500-mm length. The second part is the main experimental chamber, consisting of an octagonal stainless-steel cell in the $xy$-plane with 70-mm viewports to allow for the four 2-D MOT beams. The top part of the cell has three 19-mm OD tubes in the $xz$-plane. The two on the sides are at an angle of $\pm 15^\circ$ from the vertical, and are used for the downward-launching beams. The one in the middle ends in a region where the launched atoms can be probed, which is 22 cm above the center of the 2D-MOT. The bottom part of the cell has a large 100-mm viewport, which is large enough to accommodate the upward-launching beams. The entire chamber on this side of the differential pumping tube is pumped by a 55-l/s ion pump.

Atoms emanating from the oven are first slowed in a spin-flip Zeeman slower \cite{PAN10}, i.e.\ one with an initial field of $+290$ G and a final field of $-260$ G. The length of the slower is 0.33 m. It is designed to slow all atoms with a velocity less than 330 m/s, which corresponds to capturing 55\% of the atoms emanating from the oven when it is heated to 400$^\circ$C. The final velocity at the end of the slower is $23$ m/s. This is equal to the capture velocity of the 2D-MOT, defined as \cite{PRS10}:
\begin{equation}
v_c = \left( |\Delta| + \Gamma \right) \frac{\lambda}{2\pi} \, ,
 \label{vc}
\end{equation}
where $\Delta$ is the detuning of the beams. The coils required for generating the slower-field profile are made by winding welding wire around the outside of the vacuum system. The wire carries 34 A for the forward slower, and 26 A for the reverse part. The slowing beam has a detuning of $-420$ MHz. It is circular, with $1/e^2$ diameter of 30 mm at the entrance to the octagonal chamber, and focussed to a spot at the end of the differential-pumping tube using a convex lens of focal length 1 m.

The field gradient required for the 2D-MOT is produced by two pairs of coils, one along each set of MOT beams \cite{WAV97}. Each coil is made of 8 turns of 4-mm square copper tube, designed to have cooling water flowing inside. The coils are wound on an aluminium form of mean diameter 105 mm, and are placed 220 mm apart. With 250 A of current flowing, the field gradient in each direction is 8~G/cm. The configuration is designed so that the gradients add in the $xy$-plane and cancel along the $z$-axis. The beams have a detuning of $-30$ MHz, which is optimal for this field gradient. The total power entering the chamber along each of the two MOT-beam axes is 10 mW. The beam is elliptic with $1/e^2$ diameter of $10 \times 15$ mm, with the long axis in the $z$ direction. Therefore, the peak intensity at the beam center is 17 mW/cm$^2$. The incoming beam is circularly polarized, and is retro-reflected through a quarter-wave plate so that the return beam has the opposite polarization.

The frequency of the laser beam is monitored in a separate atomic-beam-fluorescence-spectroscopy set up. The frequency is manually adjusted to be at the fluorescence peak, and left there for the duration of the experiment. The drift of the Ti:sapphire laser is small enough that there is no significant movement away from the peak, and no noticeable change in the 2D-MOT fluorescence.

The atoms captured in the 2D-MOT are launched using the standard technique of moving molasses \cite{CSG91}. To get a detuning of $-\Gamma/2$ (which gives the lowest temperature in
one-dimensional molasses) in a frame moving towards the up beam with a velocity $v$, the detunings in the laboratory frame
should be
\begin{equation}
\Delta_{\rm up} = -\Gamma/2 - v/\lambda \ \ \ \
{\rm and \ \ \ \ }
\Delta_{\rm down} = -\Gamma/2 + v/\lambda \, .
 \label{moving}
\end{equation}
{In effect, the launch velocity is proportional to the beat frequency and to the spatial period of the moving-molasses beams. The spatial period is longer in the $z$ direction,
therefore the total launch velocity along $z$ in the presence of the two sets of launching beams is $v_z = 2v / \cos (15^\circ)$ \cite{WAV97}.} The detunings are chosen so that $v_z$ is 13.2 m/s. The launch beams are circular with $1/e^2$ diameter of 8 mm, and have a power of 6 mW each. They are linearly polarized.

The experimental parameters used in this study are summarized in Table \ref{pars}.

\begin{table}
\caption{Experimental parameters used for the continuous cold-beam experiment.}
\begin{center}
\begin{tabular}{ll}
\hline \hline
Zeeman slower beam power & 25 mW\\
Zeeman slower beam detuning  & $-420$ MHz\\
Zeeman slower length & 0.33 m \\
Velocity after slower ($v_f$) & 23 m/s\\
2D-MOT beam intensity (max) & 17 mW/cm$^2$\\
2D-MOT beam detuning & $-30$ MHz\\
2D-MOT field gradient & 16 G/cm\\
Fountain beam intensity (max) & 24 mW/cm$^2$\\
Fountain up-beam detuning & $+2$ MHz\\
Fountain down-beam detuning & $-30$ MHz\\
Designed launch velocity ($v_z$) & 13.2 m/s\\
\hline \hline
 \label{pars}
\end{tabular}
\end{center}
\end{table}

\section{Results and discussion}

\begin{figure}
\centerline{\resizebox{0.5\columnwidth}{!}{\includegraphics{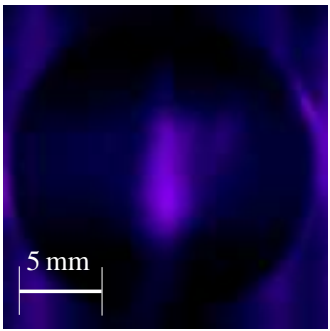}}}
\caption{(Color online) CCD image of a cold cloud of $^{174}$Yb in the 2-D MOT. There are about $7 \times 10^6$ atoms present.}
 \label{cloud}
\end{figure}

A CCD image of the cold cloud of atoms trapped in the 2D-MOT is shown in Fig.\ \ref{cloud}. The isotope used was $^{174}$Yb. The cloud appears elongated along the $z$ axis, along which there is no field gradient. This is different from the usual 3-D MOT, where the cloud looks spherically symmetric. Using a calibrated photo-multiplier tube (PMT), we estimate the number of atoms to be $7(2) \times 10^6$.

\begin{figure}
\centerline{\resizebox{0.95\columnwidth}{!}{\includegraphics{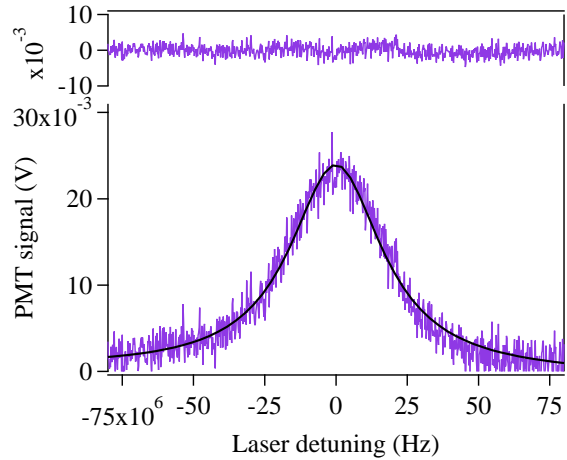}}}
\caption{(Color online) Fluorescence signal (measured using a PMT) of $^{174}$Yb atoms in the probe region, when the atoms are launched continuously from the 2D-MOT. The probe laser is scanned around the resonance peak. The solid line is a fit to a Voigt profile, with the fit residuals shown on top.}
 \label{cw}
\end{figure}

The first cold-beam experiment was designed to look at the atoms launched continuously from the 2D-MOT. The isotope used was again $^{174}$Yb. The fluorescence signal from atoms reaching the probe region was measured using a PMT (R928, Hamamatsu) by scanning the diode laser across the $^{174}$Yb resonance. The power in the diode laser was 2.5 mW in a beam of $1/e^2$ size 2.3 mm $\times$ 5.4 mm.  The resulting spectrum is shown in Fig.\ \ref{cw}. The probe laser is sent perpendicular to the launched atoms, so the expected lineshape is primarily a Doppler-free Lorentzian profile.
{There will be a Gaussian component due to the transverse spread in the beam resulting from the temperature in the 2D-MOT, but this is expected to be very small. For example, a temperature of 1 mK will result in a broadening of 0.7 MHz, which is negligible compared to the natural linewidth of 28 MHz.

This expectation is borne out by the experimental spectrum shown in the figure. The solid line is a fit to a Voigt profile---the residuals on top show that it describes the measured spectrum quite well. The Voigt shape is mainly Lorentzian with a small Gaussian component. There is some uncertainty in scaling the scan axis of the diode laser from the voltage driving the piezo-electric transducer that sets the grating angle. If we scale the axis so that the Lorentzian width is the power-broadened linewidth of 39~MHz, the width of the Gaussian component is 1~MHz. This is consistent with the temperature of 2.9(8) mK measured by us in a 3D-MOT in our earlier work \cite{PRS10}. }

The second cold-beam experiment was designed to measure the longitudinal velocity distribution of the launched atoms. For this, we look at the fluorescence signal as a function of time after the launching beams are turned on.
{Note that the 2D-MOT beams are always on.}
The probe laser is now locked to the $^{174}$Yb resonance peak, using the same spectroscopy set up as that used for the main laser. The measured signal is therefore determined by both the mean launch velocity, and the longitudinal spread around the mean. If the mean velocity is $v_z$, and the distribution follows the Maxwell-Boltzmann at a temperature $T$, then the probability density function is
\begin{equation}
f(v) = \sqrt{\frac{m}{2 \pi k_B T}} \exp \left[ -\frac{m(v-v_z)^2}{2 k_B T} \right]
\, ,
 \label{dist}
\end{equation}
so that $f(v)\,dv$ is the probability of the atom having a velocity between $v$ and $v+dv$.
The measured signal will be proportional to the total number of atoms in the probe region at a given time $t$. If the probe region is at a height $h$, then the minimum velocity that an atom must have to reach this point is $h/t + gt/2$. Therefore, the total population at time $t$ is the integral of the above velocity distribution from this velocity to $\infty$, i.e.
\begin{equation}
N(t) = N_0 \erfc \left[ \sqrt \frac{m}{2 k_B T} \left( \frac{h}{t} + \frac{gt}{2} -v_z \right) \right]
\, .
 \label{erfc}
\end{equation}
Therefore, the lineshape is essentially that of a complementary error function of $1/t$.

\begin{figure}
\centerline{\resizebox{0.95\columnwidth}{!}{\includegraphics{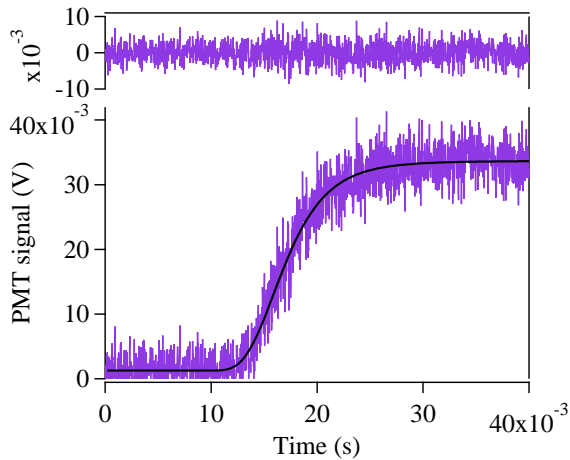}}}
\caption{(Color online) Fluorescence signal (measured using a PMT) of $^{174}$Yb atoms in the probe region as a function of time, after the launching beams are turned on at $t=0$. The probe laser is locked to the resonance peak. The solid line is a fit to the expected lineshape as given in Eq.\ \ref{erfc} in the text. The fit residuals are shown on top.}
 \label{fvst}
\end{figure}

The fluorescence signal (as measured by the PMT) in the second experiment is shown in Fig.\ \ref{fvst}. The solid line is a fit to the lineshape given above in Eq.\ (\ref{erfc}). It describes the measured spectrum very well as seen from the featureless residuals. The fit yields a mean velocity of $v_z = 13.08(4)$ m/s, consistent with the designed value. The spread around the mean is 3.4(7)~m/s, corresponding to a longitudinal temperature of 125(6)~mK.

{Interpreting this as the equilibrium longitudinal temperature for the launched atoms requires careful analysis because the atoms will take finite time to thermalize in the moving-molasses beams. And this thermalization time depends on the initial velocity distribution of the atoms. Since the 2D-MOT beams are always on, we assume that the atoms start in thermal equilibrium {\em with mean velocity equal to zero in any direction}. Then, to attain the final launch velocity of 13 m/s, they need to absorb (and emit) about 2300 photons. At the scattering rate in the moving-molasses beams, this takes 57 $\mu$s, or a distance of 0.7 mm when traveling at 13 m/s. Considering that the overlap region of the moving-molasses beams is 60 mm, it seems reasonable to expect that the atoms have reached thermal equilibrium before being launched. This also appears convincing because the mean launch velocity is close to the value calculated from the detunings of the moving-molasses beams. Therefore, we assume that the temperature measured in the second experiment is indeed the longitudinal temperature of the launched atoms.

If we take the longitudinal temperature to be 125 mK, then this is about 40 times higher than the temperature measured by us in the 3D-MOT \cite{PRS10}. It therefore seems likely that there is some heating occurring during the extraction process. Similar heating from a temperature of 45 $\mu$K in a MOT to 1 mK for the launched atoms has also been seen in the Cs experiment \cite{WAV97}. Indeed, the method of extraction seems to play a role in the longitudinal temperature, because a much smaller temperature increase (from 40 $\mu$K to 200 $\mu$K) has been reported in another Cs continuous-beam experiment where the atoms were extracted magnetically from the 2D-MOT \cite{BJD98}.}

\section{Conclusion}
In summary, we have demonstrated a continuous beam of cold Yb atoms. A continuous beam has many advantages over the more common pulsed fountain for precision measurements and optical clocks. Yb atoms emanating from a thermal source are first slowed in a Zeeman slower, then captured in a 2-dimensional magneto-optic trap, and finally launched continuously using two sets of moving-molasses beams inclined at $\pm 15^\circ$ from the vertical. About $7 \times 10^6$ atoms captured in the 2D-MOT are launched continuously with a mean vertical velocity of 13~m/s. By using a transient measurement after the launching beams are turned on, we infer a longitudinal temperature of 125(6) mK, or a velocity spread of 3.5 m/s, for the launched atoms. This is about $40 \times$ higher than the temperature in a 3D-MOT under similar conditions. The above experiments were carried out on the strongly-allowed $^1{S_0} \rightarrow {^1P}_1$ transition at 399 nm. We next plan to launch atoms by using the weakly-allowed $^1{S_0} \rightarrow {^3P}_1$  intercombination line at 556 nm.
{This should allow us to get a much lower longitudinal temperature, since the Doppler-cooling limit for this transition is only 4.4 $\mu$K, which we have shown results in a $44$ times lower temperature in the 3D-MOT \cite{PRS10}.}

\begin{acknowledgments}
The authors thank Pushpander Singh for help with the experiments. This work was supported by the Department of Science and Technology, India, through the Swarnajayanti fellowship. K.D.R. and A.K.S. acknowledge financial support from the Council of Scientific and Industrial Research, India.
\end{acknowledgments}


\end{document}